\PassOptionsToPackage{dvipsnames}{xcolor}
\documentclass[entropy,article,accept,pdftex,moreauthors]{Definitions/mdpi} 
\firstpage{1} 
\pubvolume{28}
\issuenum{1}
\articlenumber{219}
\pubyear{2026}
\copyrightyear{2026}
\externaleditor{Avishy Carmi and Eliahu Cohen} % More than 1 editor, please add `` and '' before the last editor name
\datereceived{15 December 2025} 
\daterevised{7 February 2026} % Comment out if no revised date
\dateaccepted{8 February 2026} 
\datepublished{13 February 2026} 
\doinum{10.3390/e28020219}
\usepackage{siunitx}

\usepackage{algorithm}
\usepackage{pseudo}

\usepackage{listings}
\lstdefinestyle{mylisting}{
  basicstyle=\footnotesize\ttfamily,
  breaklines=true,
  frame=single,
  framerule=0.5pt,
  columns=fullflexible,
  keepspaces=true, 
}
\usepackage[htt]{hyphenat}

\DeclareMathOperator*{\argmin}{arg\,min}
\DeclareMathOperator*{\argmax}{arg\,max}
\DeclareMathOperator*{\diag}{diag}

\usepackage{bbm}

\newcommand{\bs}[1]{\boldsymbol{#1}}

\usepackage{pgfplots, pgfplotstable}
\pgfplotsset{compat=1.18}
\usetikzlibrary{intersections, perspective, external, shapes, positioning, calc, graphs, quotes, backgrounds}
\tikzset{
  roundnode/.style={
    circle, draw=green!60, fill=green!5, very thick},
  Madrid network/.pic={
    code={%
      \node[roundnode] (N1)                     {$A_1$};
      \node[roundnode] (N2) [below right=of N1] {$B_1$};
      \node[roundnode] (N3) [below=of N2]       {$A_2$};
      \node[roundnode] (N4) [below left=of N3]  {$B_2$};
      \node[roundnode] (N5) [above left=of N4]  {$A_3$};
      \node[roundnode] (N6) [above=of N5]       {$B_3$};
}}}

\pgfplotscreateplotcyclelist{mycolorlist}{
    {MidnightBlue, thick},
    {Purple, thick, densely dashed},
    {Brown, thick, densely dashdotted},
    {RedOrange, thick, densely dotted},
    {Dandelion, thick},
    {Green, thick, densely dashed},
    {lightgray, thick, densely dashdotted}% <-- don't add a comma here
}
\usepackage{comment}

\usepackage[normalem]{ulem}
\usepackage{marginnote} % for better margin notes

\renewcommand{\hl}[1]{#1}

\Title{\hl{Switching} %MDPI: We have moved the content from the individual .tex files to the main .tex file and deleted the separate source files. Please confirm.
 Coordinator: An SDN Application for Flexible QKD~Networks} %EE: Please ensure that your intended meaning is retained in the title.
\Author{\hl{Rubén} %MDPI: Please carefully check the accuracy of names and affiliations.
 \hl{B.} %MDPI: The last name should not be abbreviated. Please confirm whether “B.” is part of the surname. If yes, please provide the complete surname in full.
 Méndez $^{1,}$*\orcidA{}, Hans H. Brunner \hl{$^{2}$}%MDPI: The numbers associated with the authors’ affiliations should appear in numerical order. Please check the suggested changes.
, Juan P. Brito \hl{$^{1,3}$}, Hamid Taramit $^{1}$\orcidB{}, Chi-Hang Fred Fung \hl{$^{2}$},\\ Antonio Pastor $^{4}$\orcidC{}, Rafael Cantó $^{4}$, Jesús Folgueira $^{1,4}$\orcidD{}, Diego \hl{R.} %MDPI: The last name should not be abbreviated. Please confirm whether “R.” is part of the surname. If yes, please provide the complete surname in full.
 Lopez $^{4}$\orcidE{}, Momtchil Peev \hl{$^{2}$}\orcidF{} and\\ Vicente Martin \hl{$^{1,3}$}}

\AuthorNames{\hl{Rubén} % Authors: Accent was added on "e".
B. Mendez, Hans H. Brunner, Juan P. Brito, Hamid Taramit, Chi-Hang Fred Fung, Antonio Pastor, Rafael Cantó, Jesús Folgueira, Diego R. Lopez, Momtchil Peev, and Vicente Martin}

\address{%
$^{1}$ \quad Center for Computational Simulation, Universidad Polit\'ecnica de Madrid, \hl{28660 Madrid}%MDPI: We add the postal code. Please confirm. The following highlights are the same.
, Spain; \hl{juanpedro.brito@upm.es (J.P.B.); hamid.taramit@upm.es (H.T.); vicente.martin@upm.es (V.M.)} %MDPI: We added these email addresses here according to those submitted online at submission system. Please confirm.
\\
\hl{$^{2}$} \quad Munich Research Center, Huawei Technologies Duesseldorf GmbH, \hl{80992 Munich}, Germany; \hl{hans.brunner@huawei.com (H.H.B.); fred.fung@huawei.com (C.-H.F.F.); momtchil.peev@huawei.com (M.P.)}\\
\hl{$^{3}$} \quad DLSIIS, ETSI Inform\'aticos, Universidad Polit\'ecnica de Madrid, \hl{28660 Madrid}, Spain\\
$^{4}$ \quad Telef\'onica gCTIO, \hl{28050 Madrid}, Spain; \hl{antonio.pastorperales@telefonica.com (A.P.); rafael.cantopalancar@telefonica.com (R.C.); diego.r.lopez@telefonica.com (D.R.L.); jesus.folgueira@telefonica.com (J.F.)}}

\corres{Correspondence: ruben.bmendez@upm.es}

\abstract{A monitor and control framework for \emph{\hl{quantum-key-distribution}%MDPI: Please confirm if the italics are necessary; if not, please remove them. The following highlights are the same.
} (QKD) networks equipped with switching capabilities was developed.
On the one hand, this framework provides real-time visibility into operational metrics.
Specifically, it extracts essential data, such as the switching capabilities of QKD modules, the number of keys stored in buffer queues of the QKD links, and the respective key generation and consumption rates along these links.
On the other hand, this framework allows \emph{\hl{software-defined networking}} (SDN) applications to operate on the collected information and address the cryptographic needs of the network.
The SDN applications dynamically adapt the configuration of the switched network to align with its changing demands, e.g.,~prioritizing key availability on critical paths, responding to link failures, or reallocating generation capacity to prevent bottlenecks.
This contribution demonstrates that the combination of switched QKD, centralized control, and global optimization strategies enables efficient, policy-driven operation of QKD networks.
The cryptographic resources are allocated to maximize performance and resilience while remaining aligned with the specific policies set by network administrators.}


\begin{document}

%\section{Introduction}
%The introduction should briefly place the study in a broad context and highlight why it is important. It should define the purpose of the work and its significance. The current state of the research field should be reviewed carefully, and key publications cited. Please highlight controversial and diverging hypotheses when necessary. Finally, briefly mention the main aim of the work and highlight the principal conclusions. As far as possible, please keep the introduction comprehensible to scientists outside your particular field of research.
\section{Introduction}\label{sec:Introduction}

The emergence of practical quantum computing represents a significant threat to the security of global communication systems.
For example, Shor’s algorithm~\cite{shor} can solve problems such as integer factorization and discrete logarithms in polynomial time.
This ability challenges the security assumptions of widely used public-key cryptographic systems, and~their compromise renders current network security protection schemes ineffective against adversaries with quantum~capabilities.

\emph{\hl{Quantum key distribution}} (QKD) is a promising technique that uses the principles of quantum mechanics to securely generate symmetric cryptographic keys between two parties~\cite{qkd}.
Unlike presently widespread classical cryptographic methods, which rely on computational hardness, QKD derives its security from the fundamental laws of quantum mechanics (QM).
As an example, in preparing and measuring QKD systems, the~no-cloning theorem guarantees that any attempt to intercept or measure a quantum signal irreversibly disturbs it. Similarly, QM principles have analogous implications for other QKD types.
For this reason, (ideal) QKD protocols are information-theoretically secure, and~thus immune to both classical and quantum computational~attacks.

Following the first experimental demonstration of QKD by Bennett~et~al. in 1992~\cite{v_bennet}, significant progress was made in developing quantum communication networks.
Early implementations such as the DARPA quantum network~\cite{darpa} and the SECOQC project~\cite{peev2009secoqc} laid the foundation for the deployment of QKD in optical network environments.
Further advances led to the creation of more expansive infrastructures, including the Tokyo QKD network~\cite{v_tokyo} and the China quantum communication network~\cite{v_china}.
The latter features satellite-based communication links to enable very-long-distance quantum key exchange.
The Madrid quantum communications infrastructure~\cite{martin2023madqci}, which has grown over the past years, focuses on interoperability and integration with heterogeneous telecommunication systems to demonstrate a feasible path toward widespread adoption of QKD in \mbox{existing~infrastructures}.

Despite these advances, QKD faces important technical and economic challenges.
The range over which quantum signals can be reliably transmitted is limited without quantum repeaters.
Standard optical amplification introduces noise, which must be attributed to the eavesdropper and may prohibit key generation in the unbounded-adversary paradigm.
For this reason, current QKD networks often rely on trusted nodes for key forwarding to extend the key distribution~range.

Additionally, as~of today, QKD requires costly and specialized hardware.
The direct connection model, in~which a transmitter (Alice) is linked to a specific receiver (Bob), limits flexibility and scalability.
To address these constraints, switched QKD architectures were proposed~\cite{quantum2030032, hansSwitch}.
In such an architecture, QKD modules can be dynamically linked to different peers depending on network demand.
This dynamic reconfiguration enables a more efficient use of the QKD modules, reduces infrastructure costs, and~improves network flexibility and scalability while preserving the inherent security properties of~QKD. 

However, coordinating QKD switching capabilities in growing network infrastructures presents non-trivial challenges.
A promising solution lies in the adoption of \emph{\hl{software-defined networking}} (SDN) \cite{Benzekki2016}, a~paradigm that enables standardized, coordinated, and~programmable network management.
SDN was introduced to reduce the vendor dependence of traditional networking architectures and has since evolved into a mature framework for orchestrating complex environments that require a high level of programmability and flexibility.
By abstracting the network status and resources to the SDN controller, SDN enables dynamic reconfiguration and real-time traffic management through the use of open~protocols.

For quantum communication networks, SDN provides the flexibility necessary to monitor complex infrastructures and coordinate efficient key-forwarding and switching strategies that adapt to network demands~\cite{sdn}.
The use of SDN allows for seamless integration of additional QKD modules into existing infrastructures and facilitates the deployment of advanced features like link prioritization and quality-of-service constraints.
SDN for QKD networks acts as an essential enabler for the transition from isolated QKD links to fully integrated and adaptive quantum secure~networks.

A novel SDN-enabled framework was developed for real-time monitoring and control of switched QKD networks.
This framework leverages the network's switching capabilities to maximize performance and resilience in accordance with switching strategies defined by the network administrator.
The main contributions of this work are listed as~follows:
\begin{itemize}
    \item The principles of switched QKD and the architecture and control of switched QKD networks are described coherently.
    \item An SDN-enabled framework is developed with two core components: (1) a formal dynamic network model that abstracts the switching capabilities of QKD modules and
    (2) a switching coordinator. Implemented as an SDN application, the~coordinator continuously manages switching among QKD modules based on real-time network state information and administrator-defined policies.
    \item The proposed framework is implemented and evaluated through simulations on an example model of a switched QKD network using two switching strategies, demonstrating efficient management of key buffers and effective prevention of key depletion.
\end{itemize}

The remainder of this paper is organized as follows. 
Section~\ref{sec:Switched QKD} details the principles of switched QKD.
Section~\ref{sec:QKD-network abstractions} defines the SDN-enabled architecture of switched QKD networks.
Section~\ref{sec:Switching coordinator} introduces the formal network model and switching coordinator in the proposed framework.
Section~\ref{sec:Optimization algorithms} presents example switching strategies and evaluates their efficiency through simulations. 
Finally, Section~\ref{sec:Conclusions} concludes the paper and outlines directions for future~work.

\section{Switched QKD~Network\label{sec:Switched QKD}}

QKD switching is a technique that allows a QKD module to connect to different peers over time.
The involved QKD modules must support this feature and be capable of rapidly synchronizing with new peers, identifying them, and automatically initiating or resuming key generation. In~general, switching can be a technique to improve the security of the network. For~example, a~trusted node can be potentially bypassed as long as the tolerable losses are low enough. This means that fewer trusted nodes are needed in the end-to-end connection~\cite{selentis2025evaluating}. It can also serve to create several disjoint paths connecting the same initial and end nodes, meaning that an attacker should gain access to all different paths to obtain the final key, making for a more robust network. It also facilitates the execution of novel protocols like quantum private query (QPQ) \cite{qpq}, creating on-demand a direct quantum channel between a user and a database. 

Throughout the text, a \emph{\hl{physical (QKD) link}} refers to a set of resources, e.g.,~QKD transmitters, QKD receivers, fibers, etc.
A physical link has to connect at least one QKD transmitter and one QKD receiver.
A \emph{\hl{fixed link}} is an always-active physical link with a dedicated set of resources.
A \emph{\hl{flexible link}} or \emph{\hl{switched link}} is a physical link that shares resources with other, mutually exclusive, physical links, e.g.,~the same QKD transmitter.
A physical link can include more than two QKD modules.
\emph{\hl{Parallel links}} are physical links that generate a key for the same node pair.
In the following, only mutually exclusive parallel links are considered\hl{. Simultaneously active parallel links are interesting in the context of vendor independence. 
The keys of two parallel QKD links, provided by different vendors, can be combined to protect against (unknown) vulnerabilities of specific products. In this contribution, simultaneously active parallel links are treated as non-parallel links of different node pairs.} %MDPI: Footnotes are not supported in our journal. We have therefore included this paragraph in the main text. Please confirm.
% Authors: We integrated the note within the text.
\emph{\hl{(QKD) link}} is used as the abstraction in the key-forwarding layer that connects a pair of module-to-module key buffers, i.e., a~link buffer (Section~\ref{sec:QKD-network abstractions}).
Such a link buffer might be filled by a \emph{\hl{set of parallel links}} or a single physical link, which can then also be unambiguously called a~link.

\emph{\hl{Configuration}} is used to refer to a set of simultaneously active physical links within the network.
The physical links that are active under the same configuration must utilize disjoint sets of resources.
Each possible configuration represents a specific association of transmitters and receivers that respects the network's switching~possibilities. 

\begin{figure}[H]
\includegraphics[width=0.61\textwidth]{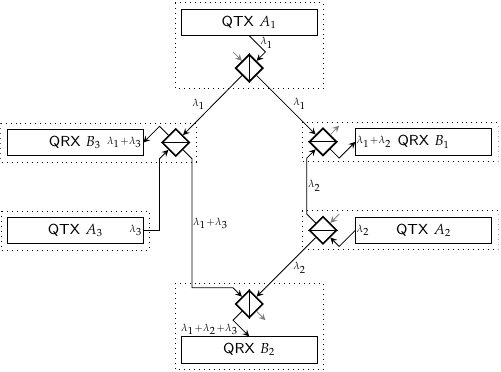}
  \caption{%
  \hl{Example} %MDPI: To avoid any changes to the images, we have replaced all images generated by the code with PDF images. Please check and confirm.
  % Authors: we confirm.
 \hl{hexagonal} %MDPI: Please confirm whether an explanation of the gray arrows and dashed boxes needs to be added to the figure caption.
 % Authors: A sentence was added to the caption.
 network with QKD modules interconnected through beam splitters (not necessarily a physically accurate depiction).
 QTX $A_{k}$ are QKD transmitters, and QRX $B_{l}$ are QKD receivers, $\lambda_{k}$ is the wavelength of the QKD signal of QTX $A_k$.
  Each receiver can see signals from multiple transmitters, represented with the plus symbol.
  By tuning the reception wavelength, a receiver can decide which transmitted signal it is detecting.
  \hl{Light gray arrows indicate unused ports of a beam splitter, dotted boxes the perimeter of a node.}%
  \label{fig:splitter network}}
\end{figure}

\begin{figure}[H]
\includegraphics[width=0.3\textwidth]{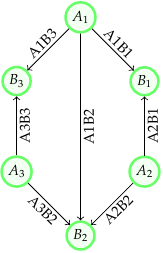}
   \caption{Graph derived from the network in Figure~\ref{fig:splitter network} with abstract QKD links on the key forwarding layer.
   The nodes are transmitters or receivers, while the edges represent the abstract links, directed from a transmitter to a receiver.
   \label{fig:graph network}}
\end{figure}

In this work, the~example network illustrated in Figure~\ref{fig:splitter network} is considered.
This hexagonal network, equipped with switching capabilities, serves as the basis for demonstrating the proposed design and framework throughout the paper.
Nevertheless, the~proposals presented in this work apply to any topology of a switched QKD network.
The abstract links of the example network are shown in Figure~\ref{fig:graph network}.%, while the physical links and some configurations are included in \hl{Figure}%MDPI: The first occurrence of each figure citation should be in numerical order. “Figure 5” appears after “Figure 2”. Please modify.
%~\ref{fig:useful configurations}.
%Authors: We updated the text accordingly. 

\vspace{-9pt}

In a flexible QKD network, it might be necessary to switch between multiple configurations to allow key generation on all links, e.g., switching between the \hl{configurations} %MDPI: Please ensure all variables/values in the equation appear in the same format in the text (normal/italic/bold/subscript/superscript).
%Authors: Done.
\begin{eqnarray}
  \mathcal{C}_1 & = & \{\{A_1,B_1,\lambda_1\}, \{A_2,B_2,\lambda_2\}, \{A_3,B_3,\lambda_3\}\}\notag\\
  \mathcal{C}_2 & = & \{\{A_1,B_3,\lambda_1\}, \{A_2,B_1,\lambda_2\}, \{A_3,B_2,\lambda_3\}\} \label{eqn:simple configs}
\end{eqnarray}
in the network of Figure~\ref{fig:splitter network}, where each link is represented by a transmitter-receiver pair and a wavelength.
The evolution of a switched QKD network over time can, therefore, be viewed as a sequence of configurations, each maintained active for a certain~duration.

Three major motivators were identified that promote breaking up fixed QKD links and deploying switched QKD networks.
The first motivation is the ability to recover from faulty links by switching to a backup line within a set of parallel links.
This is also relevant for avoiding downtime during maintenance.
Typically, this would be realized with optical switches to redirect the QKD signal over a different~fiber.

The second motivation is a reduction in installation cost.
A fully-connected, fixed QKD network of $N$ nodes requires $\left(N^2-N\right)$ QKD modules (two per edge), whereas a switched QKD network can, in~principle, reduce this to $N$ QKD modules (one per node).
In~\cite{hansSwitch}, 17 QKD links were realized with only 5 transmitters and 5 receivers.
In contrast, 17 pairs would have been necessary with fixed QKD links.
In trusted-node line networks, common for QKD backbone networks \cite{v_china, horoschenkoff2025demoquandtcarriergradeqkdnetwork}, the~number of required modules can almost be halved from two per hop ($2(N-1)$) to one per node ($N$).
However, each transmitter and each receiver can only be active for a single link at a time.
Therefore, because there might be other shared resources in the network~\cite{hansSwitch}, switchable QKD links can be mutually exclusive.
In this case, switching is required during normal operation to allow keys to be generated on all~links.

The third motivation is the optimization of key generation by adaptively allocating resources according to changing network needs~\cite{kanellos}.
By coordinating the link switching, networks can efficiently manage limited resources and maintain robust, secure communications even in dynamic or failure-prone environments.
Moreover, the~ability to anticipate future demand and proactively adjust key generation positions in SDN QKD networks as highly responsive~systems.

Depending on the capabilities of the QKD modules and the type of switching, connecting new peers can introduce significant downtime.
Reconfigurations of hardware properties can easily take several seconds, e.g.,~for settling times and initial synchronization, while software-based changes can be quick.
The key generation delay with large block sizes can also cause severe downtime for newly established links.
Therefore, switching should occur at a sufficiently low frequency to avoid depleting key buffers; otherwise, switching delays may degrade overall network~performance.

Several different switching techniques were developed to dynamically control which peers' keys are generated in the network.
Three possible approaches for QKD networks are summarized in the~following.

\subsection{Optical Path~Switching}
Optical switches are used to dynamically reconfigure the physical routes connecting QKD transmitters and receivers.
Rather than maintaining fixed fiber connections between all node pairs, optical path switching allows, e.g.,~a single transmitter to sequentially establish point-to-point links with multiple receivers.
Optical switches are well-established network devices and can have low insertion loss with minor impact on the key generation capabilities.
The switches can be extended with multiplexers to form wavelength-selective switches, but they must not contain amplifiers as is typical in commercial \emph{\hl{reconfigurable optical add-drop multiplexers}} (ROADMs) \cite{8884156}.

\subsection{Wavelength~Switching}
Wavelength switching refers to changing the optical frequency (or channel) used for QKD within a \emph{\hl{wavelength division multiplexing}} (WDM) framework.
Combined with WDM equipment, e.g.,~ROADMs, wavelength switching extends the optical path switching capabilities of a network, and can increase the availability of paths to avoid congestion by adding parallel links~\cite{hansSwitch}.

In network segments that are equipped with optical beam splitters rather than switches, wavelength switching can also be used to create different pairings between the QKD modules.
For example, a~continuous-variable QKD receiver that sees signals from different transmitters, each on its own wavelength, through beam splitters, can tune the detection wavelength to select a transmitter with which it then generates a key.
Figure~\ref{fig:splitter network} depicts an example network with QKD modules interconnected in a hexagon through beam splitters.
Two straightforward configurations of this network would be that each receiver tunes its wavelength to the neighboring transmitter in a counter-clockwise or clockwise direction ($\mathcal{C}_1$ and $\mathcal{C}_2$ in \mbox{Equation \eqref{eqn:simple configs})}, respectively.

\subsection{Protocol~Switching\label{subsec:Protocol switching}}

A more abstract way of switching is protocol switching, which can be useful for multiple reasons.
One possibility is emerging multi-party protocols~\cite{hajomer2024}.
For example, in~a network segment with beam splitters, a~QKD transmitter generates a key with multiple receivers at the same time, where the involved modules agree on a conference key \hl{(this requires generalizing the notation from node pairs to node groups), }%MDPI: Footnotes are not supported in our journal. We have therefore included this paragraph in the main text. Please confirm. 
% Authors: We integrated the note within the text.
or where some of the receivers support and boost the key generation of another receiver.
Each of these possibilities constitutes a different protocol, as well as a different, possibly parallel, physical link.
From the network control point of view, this gives rise to multiple mutually exclusive physical links that can be coordinated to the benefit of the~network.

In the example network in Figure~\ref{fig:splitter network}, all receivers could tune their wavelength to detect the QKD signal of transmitter $A_1$.
Then, this transmitter could generate a key with one receiver after another, while the other receivers support the key generation.
The different combinations of transmitter $A_1$ generating a key with receiver $B_1$, with or without the support of $B_2$ and/or $B_3$, form a set of parallel links between the node pair of $A_1$ and $B_1$.

Another reason for protocol switching is that a QKD module might support different protocols, based on different security assumptions, for~different security levels.
Since key rates with strict security assumptions are often limited, there might be an interest in running protocols with loosened security assumptions to increase the key rate for less critical communication.
Typical ideas for loosening security assumptions are trusting parts of the channel or bounding the capabilities of a potential~attacker.

\section{QKD-Network~Abstractions\label{sec:QKD-network abstractions}}

\begin{figure}[H]
\includegraphics[width=0.61\textwidth]{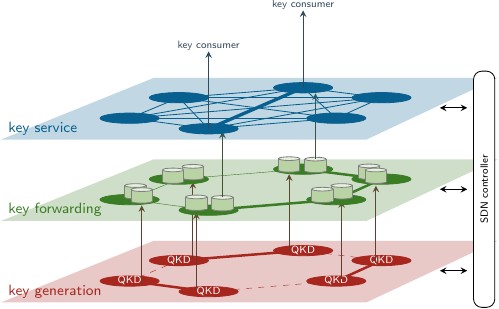}
   \caption{
   \hl{QKD-network} %MDPI: 1. We moved the figure after where it's first mentioned in the main text. Please confirm. 2. Please confirm whether an explanation of the arrows and dashed lines needs to be added to the figure caption.
   % Authors: 1. We confirm. 2. No explanation needed.
 abstractions: The key-service layer presents end-to-end keys (bold line) to key consumers; it handles access control and user prioritization.
   These end-to-end keys are generated with low delay on the key-forwarding layer based on keys taken from module-to-module link buffers along a key-forwarding route (bold line).
   The link buffers are continuously being filled by active physical links (bold lines) on the key generation layer.
   The set of active physical links changes on a comparatively slow time scale.%
   \label{fig:QKD-network abstractions}}
\end{figure}

To establish the structural foundation of the proposed framework, a~three-layer architecture for flexible QKD networks is introduced, as~illustrated in Figure~\ref{fig:QKD-network abstractions}.
Each layer fulfills distinct responsibilities in the generation, distribution, and~consumption of quantum keys.
Central to this architecture is the notion of link buffers, designated memory that temporarily holds quantum keys output by QKD modules for a specific node pair, a~single physical link, or a set of parallel links (see Section~\ref{sec:Switched QKD}).
These link buffers act as an intermediary between the generation of direct module-to-module keys and their usage in the~network.

A flexible QKD network comprises a range of interconnected entities beyond QKD transmitters and receivers: switching modules, link buffers, key forwarding modules, key managers, and cryptographic applications.
In the following sections, we examine each layer of the QKD-network architecture, highlighting the specific components involved, their responsibilities within the network, and~their role in enabling programmable, optimized quantum key distribution.
This analysis not only clarifies the operational structure of the\linebreak system, but also lays the groundwork for understanding how SDN applications---running on the SDN controller---can leverage real-time information from each layer to make informed, goal-oriented decisions.

\subsection{Key~Service}

The top-level abstraction is the local key-store interface to end-user applications that rely on cryptographic keys for secure communication.
This layer hides the complexity of key forwarding and generation and provides end-to-end keys (on demand or streamed) by forwarding key requests to the underlying layers.
There might be application- or peer-node-specific link buffers in this layer to reduce key-request latencies.
This layer manages all topics related to user access and prioritization and imposes the demand that drives the behavior of the rest of the network.
Example key consumers that connect to this layer are conventional cryptographic applications, such as transport layer security (TLS) services, internet protocol security (IPsec), or~cryptographic devices, such as~encryptors.

\subsection{Key~Forwarding}

This layer hosts a link buffer for every potential module-to-module physical QKD link or set of parallel links.
By operating on the link buffers, which are being filled from the underlying key generation layer, the~forwarding layer becomes independent of QKD itself.
As is typical for trusted node architectures, the~responsibility of this layer is to route keys between nodes across the QKD network to enable end-to-end key distribution on demand with low latency.
This layer plays a vital role in balancing key consumption between different link buffers and performing retransmissions if needed.
Techniques to optimize key delivery paths---such as those proposed in~\cite{piotrFK}---are critical to ensure that forwarding delays or inefficiencies do not result in key depletion.
Such optimization techniques can be formalized as SDN applications, which are outside the scope of this~work.

\subsection{Key~Generation}

At the base of the architecture is the physical generation of module-to-module QKD keys.
An active physical QKD link continuously pushes key material into the associated link buffer on the forwarding layer. \hl{Because of their real-time nature and efficient implementations, active physical QKD links continuously generate key material.
Architectures exist that model on-demand key generation by the QKD modules.
In this case, the~link buffers are inside the QKD modules, which is not in conflict with the given abstractions.}
%MDPI: Footnotes are not supported in our journal. We have therefore included this paragraph in the main text. Please confirm.
% Authors: We integrated the note within the text.
While fixed point-to-point QKD pairs are links that might always be active, not all links might be active at the same time in QKD networks that include switching~capabilities.

The link buffers are critical for decoupling key-consumer demands with low delay requirements from the physical generation process, which involves slow switching between different sets of active physical links.
Each link buffer has a finite capacity and is, therefore, subject to underflow (key depletion) or overflow (wasted generation), both of which can degrade network efficiency.
Maintaining an appropriate balance across these distributed link buffers must be taken into account to succeed in fulfilling the demands of the network.
The process of deciding the active set of physical links is one of the main contributions of this work, discussed from Section~\ref{sec:Switching coordinator} onward.
Although there might be a benefit to joint optimization of key-forwarding routes and the selection of active physical links, these tasks are deliberately divided to manage the network~optimization.

\subsection{SDN~Control}

The switching techniques and network abstractions described in the previous sections are the foundation for a flexible and programmable QKD network.
However, the~effectiveness of such an infrastructure highly depends on a decision-making entity capable of coordinating these components according to real-time conditions and operational goals.
This is particularly critical in environments where resources are limited, key demand fluctuates dynamically, and~uninterrupted key availability is~essential.

\begin{figure}[H]
\includegraphics[width=0.435\textwidth]{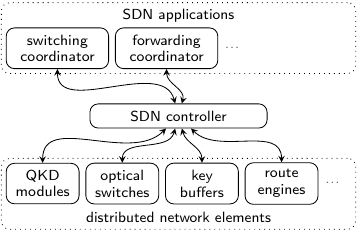}
   \caption{%
   \hl{SDN} %MDPI: We moved the figure after where it's first mentioned in the main text. Please confirm.
   % Authors: We confirm.
 adaptation for QKD networks.
   The logically centralized SDN controller provides network views of the switching and key-forwarding capabilities of the distributed network elements to the SDN applications.
   According to a given policy, the~SDN applications, e.g.,~the switching coordinator, provide decisions on how the SDN controller should configure the network elements.
   The communication between the different network elements within a (trusted) node and the SDN controller is typically channeled through an SDN agent in the node.
   \label{fig:SDN adaptation for QKD networks}}
\end{figure}

This decision-making role is realized by the SDN controller.
The adaptation of SDN for QKD enables programmable and adaptive control of quantum key resources through a logically centralized SDN controller (Figure~\ref{fig:SDN adaptation for QKD networks}).
On the one hand, the~SDN controller derives abstract views of the network from the advertised capabilities of the network elements and provides these views to the SDN applications. On~the other hand, it implements the decisions of the SDN applications regarding the desired network behavior on the network elements~\cite{Benzekki2016}.
Network elements are physical or logical network devices that advertise their capabilities and allow control over them.
In addition, the~SDN controller can provide event notifications to the SDN~applications.

A typical example of an SDN approach in classical networks is the use of OpenFlow~\cite{openflow}, where a centralized controller programs the routing tables of Ethernet switches.
This illustrates the core SDN principle: separating the control plane (where routing decisions are made) from the data plane (where packets are forwarded).
The same paradigm can be applied to QKD networks.
For instance, SDN can be used to manage dynamic switching of active quantum links.
Separately, SDN can also control key forwarding by deciding and configuring the forwarding routes~\cite{sdn}. While SDN provides flexible mechanisms for managing authentication procedures, QKD protocols require information-theoretically secure authentication, typically established through a pre-shared key during device installation or via an initial short-lived public key~\cite{wang2021experimental}. Although~this initial authentication phase may benefit from SDN support, it is outside the scope of this~work.

SDN applications are programs that run on the SDN controller.
Based on the provided network views, SDN applications provide instructions for configuring the network according to given policies.
This modular approach splits distinct control tasks from the SDN controller's core logic into dedicated SDN applications.
Consequently, this architecture greatly simplifies the implementation and maintenance of control functions, as it abstracts the complexities of interacting with vendor-specific devices and protocols.
An SDN application delegates all communication with the network elements to the SDN controller,
which in turn executes commands through standardized management protocols such as NETCONF~\cite{enns2011rfc} and RESTCONF~\cite{rfc8040Restconf}.
This modularity promotes portability, reusability, and~ease of integration, enabling SDN applications to operate across heterogeneous infrastructures without requiring internal adaptations for new hardware components.
Additionally, this architecture allows flexible integration of high-level orchestration schemes without modifying the interface between the network elements and the SDN controller.
In the next section, an~SDN application for coordinating switching in QKD networks---one that decides which physical links are active on the key generation layer---is investigated.

\section{Switching~Coordinator}\label{sec:Switching coordinator}

The switching coordinator is implemented as an SDN application.
In the SDN paradigm, SDN applications communicate with the SDN controller to both retrieve information about the current state of the network and instruct it on the desired operational behavior.
Based on collected data about the network status, the~switching coordinator conceptually operates as an interrupt-driven state machine that can run optimization algorithms to continuously determine the optimal configuration of the network.
When a new configuration, i.e.,~the next set of physical links that should be active in the QKD network, needs to be applied, the~switching coordinator interacts with the SDN controller in order to reconfigure the QKD network and steer it toward the desired~behavior.

From the SDN controller, the switching coordinator retrieves the current state and structure of the network. 
The structural information contains the list of physical QKD links that are available through switching and their associated key generation rates, which are assumed to be predicted from static channel conditions.
Accordingly, the~current structure of the network is updated by the SDN controller whenever there is a change in the network elements.
The current network state consists of real-time data comprising the filling levels of the link buffers and~their output rates, i.e.,~the rates at which they are being depleted.
Here, it is assumed that the key generation rate of each physical link is constant and known.
Varying key generation rates can be included in the real-time data.
Both the network structure and the network state are formalized in the input data model described in Section~\ref{subsec:Formal network model}.

Once the relevant data are collected, the~switching coordinator invokes one or more optimization algorithms.
Based on the policies and objectives defined by the network administrator and the current availability of resources, these algorithms compute configuration-duration tuples (Section~\ref{subsec:Periodic and event-driven operation}).
Each configuration is a set of physical links that are active in the QKD network at the same time.
Examples of such optimization algorithms that determine which physical links to activate are described in detail in Section~\ref{sec:Optimization algorithms}.
These algorithms are designed to cover the needs that the network may have, the most common objective being to avoid key depletion, although other requirements defined by the network administrator can overrule this~objective.

These evolving decisions on which configuration to use are the basis of the network operation.
It is assumed that the SDN controller has full knowledge of all network elements, e.g.,~the QKD modules and switches, their capabilities, and their interconnections.
This enables the SDN controller to maintain a full picture of the network structure and~hence provide a list of all potential physical links.
Based on this list, a set of feasible configurations is generated, where each configuration consists of the physical links it activates along with the associated key generation rates.
The operation of the switching coordinator is based on the~following:
\begin{itemize}
    \item Network structure: infrastructure data formalized in structured lists characterizing the network elements and their configurable capabilities.
    \item Network state: operational data provided as real-time states of the buffers across all links, along with the current rates of key consumption.
    \item Optimization algorithms: provided as switching strategies or objectives specified by the network administrator.
\end{itemize}
When the switching coordinator decides to reconfigure the network into a new configuration, it issues the necessary instructions to the SDN controller in order to activate the new set of physical links.
This procedure depends on the available types of switching as defined in Section~\ref{sec:Switched QKD}.

\subsection{Formal Network~Model\label{subsec:Formal network model}}

A formal network model is built upon static infrastructure data and dynamic operational data.
The infrastructure data defines the network structure, consisting of attributes characterizing the network elements, their capabilities, and~their interconnections.
The operational data constructs the network state, which comprises real-time information on the states of all links across the network.
This separation between static and dynamic data acquisition allows the system to minimize overhead, focusing computational effort on updating only the essential information needed to support adaptive optimization algorithms without rebuilding the entire network model unless~necessary.

\subsubsection{Network~Structure}

The SDN controller is assumed to continuously maintain full knowledge of the network elements, their interconnections, and~the different switching capabilities.
That is, whenever an element or a connection is added, removed, changed, fails, or~is restored, the~SDN controller is notified.
The infrastructure data are then formalized as up-to-date, structured lists, maintained by the SDN controller.
The generation of these lists is performed during the initial discovery phase and upon the occurrence of a relevant topology-\linebreak changing event.

Here, the~objective of formalizing the list characterizing the network infrastructure is to generate the set of feasible configurations from which the switching coordinator selects.
Additionally, these lists provide a framework for translating the decisions derived by the switching coordinator into constructive configuration-deployment instructions.
Hereafter, based on information about the network elements and their interconnections, two lists are formalized: the list of links and~the list of~configurations.

To ensure compatibility with current standardization efforts and promote interoperable implementations,
the parameters extracted from QKD modules are aligned with the QKD interface definitions provided in the ETSI GS QKD 015 standard~\cite{etsigsqkd015}.
A particular focus is placed on the information that describes the operational capabilities of the QKD device, such as its role (transmitter or receiver), supported wavelengths, and supported protocols.
These parameters are consolidated into the list of network elements, serving as a foundation to construct the list of links and the list of configurations.
The attributes in the list of network elements are extracted by the SDN controller using NETCONF~\cite{enns2011rfc} or RESTCONF~\cite{rfc8040Restconf} protocols, which provide access to the configuration and status of network components in a structured and vendor-agnostic format.
The interconnections might be manually added and~maintained.

In addition to device-specific attributes, the~ability of a QKD device to dynamically interact with multiple peers---enabled by switching (Section~\ref{sec:Switched QKD})---must also be modeled.
The interconnection information and the attributes in the list of network elements define a set of interconnection rules that are crucial in determining which combinations of devices can coexist in a valid network configuration. These rules serve as a filter and foundation for generating the subsequent list of links, which captures the set of all feasible QKD channels that can be established given the device capabilities and~constraints.

Whenever the list of network elements is updated, the~list of abstract links is subsequently generated by computing all possible physical links.
A physical link is defined as a set of resources connecting at least a QKD transmitter and a QKD receiver and is feasible under acceptable operating conditions.
An abstract link is defined as a set of mutually exclusive physical links that fill the same link buffer, e.g.,~key generation with and without the help of additional~receivers. 

The following describes the attributes within the list of~links:
\begin{enumerate}[leftmargin=*]
    \item \texttt{\hl{link\_id}%MDPI: Please check the paper and confirm whether to keep the different font.
    % Authors: We confirm that we want to keep the different font.
}: Unique identifier of the link (link buffer)\hl{. A link buffer is a pair (or group) of distributed buffers with the promise of identical content. Keys can only be written to or taken from all the buffers of a link buffer simultaneously.}%MDPI: Footnotes are not supported in our journal. We have therefore included this paragraph in the main text. Please confirm.
        % Authors: We integrated the note within the text.
    \item \texttt{buffer\_capacity}: Maximal storage capacity of the link buffer.
    \item \texttt{physical\_links}: Set of mutually exclusive parallel links that can be used to fill the link buffer, comprising:
    \begin{enumerate}[leftmargin=*]
        \item \texttt{physical\_link\_id}: Unique identifier of the specific physical link.
        \item \texttt{resources}: List of resources required for the specific physical link. This connects at least one QKD transmitter and one QKD receiver that can fill the link buffer. This can also contain additional resources, e.g.,~receivers used to boost the key generation of the link.
        \item \texttt{generation\_rate}: Key generation rate of the specific physical link.
            \hl{The generation rate of an active physical link is reported by the QKD devices and can also be read from the associated link buffer itself. The last known generation rate of inactive physical links can be used as the generation rate if no other means, e.g., link monitoring, are available.}%MDPI: Footnotes are not supported in our journal. We have therefore included this paragraph in the main text. Please confirm.
            % Authors: We integrated the note within the text.
    \end{enumerate}
\end{enumerate}

\begin{figure}[H]
\includegraphics[width=\textwidth]{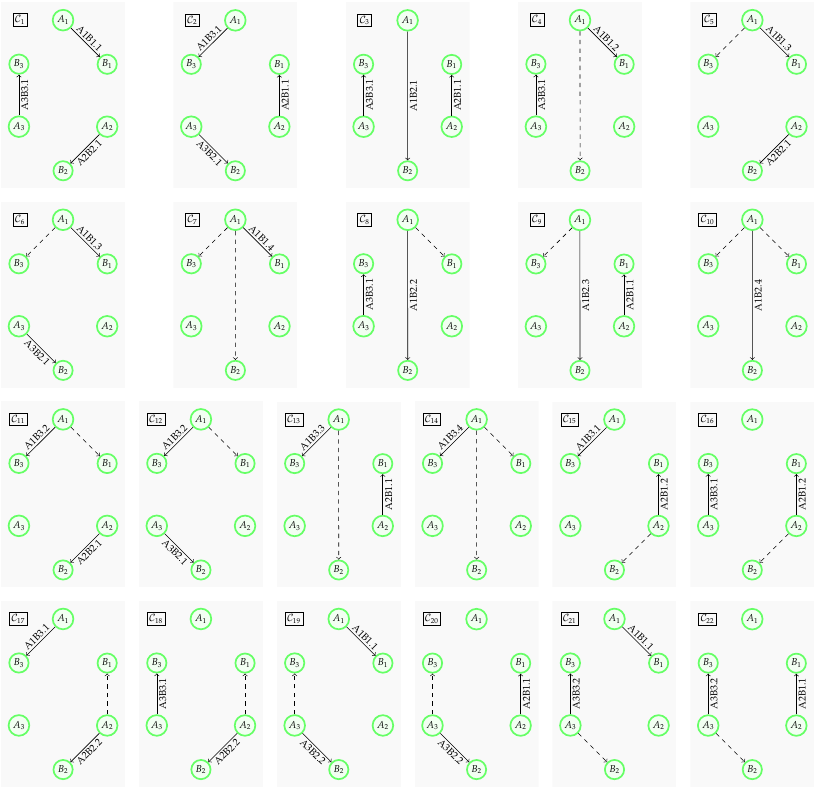}
    \caption{\hl{The} %MDPI: 1. Figures are usually put after where they're first mentioned in the main text. But the first citation of Figure 5 appears after Figure 2. Please modify. 2. Please confirm whether an explanation of the A1, A2, A3, B1, B2, B3, and dashed arrows needs to be added to the figure caption.
    %Authors: 1. We fixed that issue. 2. No explanation needed. 
 22 \emph{\hl{useful}} network configurations of the hexagonal network from Figure~\ref{fig:splitter network}.
    Configurations $\mathcal{C}_1$, $\mathcal{C}_2$, and~$\mathcal{C}_3$ contain three physical links each.
    While these links are without supporting receivers, the remaining configurations contain a physical link that is boosted by one or two supporting receivers (dashed lines).
    Any other configuration is either a subset of a given configuration, e.g.,~$\{\mathrm{A1B1.1}\}\subset\mathcal{C}_1$,
    or the physical links in the configuration are subsets of the physical links of a given configuration, e.g.,~$\mathrm{A1B1.1}\subset\mathrm{A1B1.3}$ when comparing $\{\mathrm{A1B1.1}, \mathrm{A3B2.1}\}$ and $\mathcal{C}_6$.
    \label{fig:useful configurations}}
\end{figure}

An excerpt of the list of links of the example network in Figure~\ref{fig:splitter network} is shown in Listing~\ref{lst:links} with the structural part of the proposed model only.
The abstract links can also be seen in Figure~\ref{fig:graph network}, while the physical links are included in Figure~\ref{fig:useful configurations}.
Transmitter $A_1$ can reach each receiver; there is one abstract link per receiver, each with four physical links.
The physical links differ in the set of additional receivers used for boosting the key generation.
Transmitters $A_2$ and $A_3$ can only reach two receivers each, which also limits the number of physical links to 
two per transmitter-receiver pair, with or without the support of the second~receiver.

From the previously defined set of physical links, an~SDN application derives the set of valid network configurations. Each configuration corresponds to a subset of physical links that can operate simultaneously without causing interference, resource contention, or violation of hardware or policy constraints. 
For instance, a~QKD transmitter is restricted to serving only one receiver at a time, thereby introducing a one-to-many \mbox{scheduling~limitation}.

\begin{listing}[H]
\caption{\hl{Excerpt} %MDPI: We set the formats of all listings. Please check and confirm.
%Authors: We confirm.
 of the list of links of the network in Figure~\ref{fig:splitter network}.}
\label{lst:links}
\rule{\columnwidth}{1pt}
\raggedright\footnotesize
\begin{lstlisting}[frame=none, basicstyle=\footnotesize\ttfamily, breaklines=true, columns=fullflexible, keepspaces=true]
link_id: A1B1                   # ID of the Link
buffer_capacity: 3.6MB          # Current capacity of the buffer
physical_links:                 # List of physical links
  - physical_link_id: A1B1.1    # Specific ID Link
    resources: [A1, B1, $\lambda_1$]     # Resources of this specific link 
    generation_rate: 9.6kbps    # Generation rate of the link in kbits/s 
  - physical_link_id: A1B1.2    ...
    resources: [A1, B1, B2, $\lambda_1$] ...
    generation_rate: 11.2kbps   ...
  - physical_link_id: A1B1.3    ...
    resources: [A1, B1, B3, $\lambda_1$] ...
    generation_rate: 11.2kbps   ...
  - physical_link_id: A1B1.4    ...
    resources: [A1, B1, B2, B3, $\lambda_1$] ...
    generation_rate: 12.8kbps   ...
\end{lstlisting}
\rule{\columnwidth}{1pt}
\end{listing}

The resulting list of configurations defines all admissible operational states of the QKD network. These configurations form the discrete decision space from which the switching coordinator selects the optimal sequence of actions to guide the evolution of the network in alignment with the predefined~objectives.

The set of configurations can be generated from the list of physical links in a combinatorial way.
Fixed links can be ignored in the compilation of configurations, as they are not relevant for the optimization.
These configurations represent the actionable decision space for the switching coordinator and~serve as input for the optimization algorithms introduced in Section~\ref{sec:Optimization algorithms}.

It is worthwhile to focus on \emph{\hl{useful}} configurations, a~reduced list of configurations, which does not contain configurations that are superseded by other configurations.
A configuration supersedes another configuration if~it generates at least as many keys for each link.
For example, a~configuration that is a subset of another configuration is superseded.
The configuration of only activating link ${A_1, B_1}$ in the setup of Figure~\ref{fig:splitter network} is superseded by configuration $\mathcal{C}_1$ from Equation~\eqref{eqn:simple configs}.
All useful configurations of the network in Figure~\ref{fig:splitter network} are presented in Figure~\ref{fig:useful configurations}. The~framework is designed to take as input a generic list of links and configurations. Enumerating all links and configurations is a combinatorial problem that scales in a prohibitive manner for growing numbers of nodes. The~generation of the input of these lists is out of the scope of this work. Here, it is assumed to be given. This topic and how to overcome the scaling problem will be addressed in follow-up~publications.

Listing~\ref{lst:configs} shows a portion of the list of configurations for the network in Figure~\ref{fig:splitter network}, namely the configurations $\mathcal{C}_1$, $\mathcal{C}_4$, $\mathcal{C}_6$, and~$\mathcal{C}_7$.
Each configuration is defined by a unique identifier (\texttt{config\_id}) and a set (\texttt{active\_links}) of links that are activated under that configuration. 
An active link is characterized by the attributes \texttt{link\_id}, \texttt{physical\_link\_id}, and~\texttt{key\_generation\_rate}, which are retrieved from the list of physical~links.

\begin{listing}[H]
\caption{Excerpt of the list of configurations for the network in Figure~\ref{fig:splitter network}.}
\label{lst:configs}
\rule{\columnwidth}{1pt}
\raggedright\footnotesize
\begin{lstlisting}[frame=none, basicstyle=\footnotesize\ttfamily, breaklines=true, columns=fullflexible, keepspaces=true]
- config_id: C1
  active_links:
  - link_id: A1B1
    physical_link_id: A1B1.1
    generation_rate: g(A1B1.1)
  - link_id: A2B2
    physical_link_id: A2B2.1
    generation_rate: g(A2B2.1)
  - link_id: A3B3
    physical_link_id: A3B3.1
    generation_rate: g(A3B3.1)
- config_id: C4
  active_links:
  - link_id: A1B1
    physical_link_id: A1B1.2
    generation_rate: g(A1B1.2)
  - link_id: A3B3
    physical_link_id: A3B3.1
    generation_rate: g(A3B3.1)
- config_id: C6
  active_links:
  - link_id: A1B1
    physical_link_id: A1B1.3
    generation_rate: g(A1B1.3)
  - link_id: A3B2
    physical_link_id: A3B2.1
    generation_rate: g(A3B2.1)
- config_id: C7
  active_links:
  - link_id: A1B1
    physical_link_id: A1B1.4
    generation_rate: g(A1B1.4)
\end{lstlisting}
\rule{\columnwidth}{1pt}
\end{listing}

\subsubsection{Network~State}

The current network state is a set of operational data that reflects the instantaneous states of all links across the network.
It is acquired by the switching coordinator from the SDN controller whenever it is triggered to make switching decisions.
Listing~\ref{lst:current_state} shows how the states of the links are formalized, depicting link A1B1 from the network in Figure~\ref{fig:splitter network}.
The state of each link is defined by the link identifier (i.e., \texttt{link\_id}), the~current buffer fill level (\texttt{buffer\_fill\_level}), and~the current key consumption rate (\texttt{consumption\_rate}).

\begin{listing}[H]
\caption{Excerpt of operational data for the network in Figure~\ref{fig:splitter network}.}
\label{lst:current_state}
\rule{\columnwidth}{1pt}
\raggedright\footnotesize
\begin{lstlisting}[frame=none, basicstyle=\footnotesize\ttfamily, breaklines=true, columns=fullflexible, keepspaces=true]
link_id: A1B1
current_state:
  buffer_fill_level: 3MB      # Amount of key material in this link buffer. 
  consumption_rate: 3.49kbps  # Consumption rate in kbit/s
\end{lstlisting}
\rule{\columnwidth}{1pt}
\end{listing}

\subsection{Workflow of the Switching~Coordinator}

The network model is fully constructed within the SDN controller and is made accessible to the switching coordinator.
The switching coordinator is responsible for continuously maintaining the switching configuration under dynamic network conditions.
It retrieves the full network model on demand from the SDN controller and~ensures that switching decisions are based on real-time information on the network status.
Switching operations may be initiated periodically, in~response to specific events, or through a hybrid scheme that combines both~approaches.

\subsubsection{Periodic and Event-Driven~Operation\label{subsec:Periodic and event-driven operation}}

\begin{figure}[H]
\includegraphics[width=0.575\textwidth]{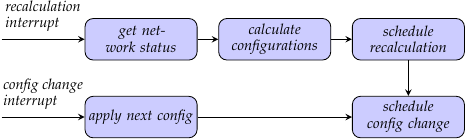}
   \caption{\looseness=-1%
   Interrupt-driven flowchart of the switching coordinator.
   The switching coordinator has two main loops, the~recalculation loop and the configuration-change loop.
   The recalculation loop is triggered by a notification from the SDN controller or according to a schedule.
   In this loop, configuration-duration tuples are calculated according to the status of the network.
   The configuration-change loop cycles through the list of configuration-duration tuples and reconfigures the network accordingly.
   \label{fig:Interrupt driven flowchart of switching coordinator}}
\end{figure}

In the case of periodic operation, the~switching coordinator respects the potential necessity of switching by acting on a repeated list of configuration-duration tuples, which it applies in a sequence.
Figure~\ref{fig:Interrupt driven flowchart of switching coordinator} depicts, on the lower row, an interrupt-driven flowchart for the continuous reconfiguration of the network according to such a schedule.
When a \emph{\hl{config change interrupt}} is raised, the~next configuration from the list is applied in the network, and the next \emph{\hl{config change interrupt}} is scheduled according to the associated~duration.

For dynamic networks with changing demand, the switching coordinator recalculates the list of configuration-duration tuples periodically to be able to adapt to the changing requirements.
In networks with constant generation and consumption, the~list of configuration-duration tuples only needs to be computed once as long as nothing happens.
The periodic recalculation based on interrupts is shown in the top row of Figure~\ref{fig:Interrupt driven flowchart of switching coordinator}.
When a \emph{\hl{recalculation interrupt}} is raised, the~SDN controller is queried for the current network status, and a new list of configuration-duration tuples is calculated.
Subsequently, the~next \emph{\hl{recalculation interrupt}} is scheduled according to the recalculation period, as well as an immediate \emph{\hl{config change interrupt}}.

In the case of event-based operation, the~switching coordinator waits for an event notification from the SDN controller to trigger a \emph{\hl{recalculation interrupt}}.
Relevant events are, e.g.,~the detection of a failing link, the detection of new equipment, the~filling of a link buffer reaching a lower threshold, or~a request by the network operator.
Event-based recalculation can be seen as a special case of the periodic recalculation, where a single configuration with infinite duration (no interrupt scheduled) is selected.

Event-based recalculation can be combined with periodic recalculation, with~the period defined according to the network requirements.
Periodic recalculation may serve as the primary approach, while event-based recalculation is used as a safeguard to address critical network conditions.
This hybrid approach can be further improved by learning and modeling the behavior of network conditions.

\subsubsection{Switching~Decisions}

When the switching coordinator is triggered to provide switching decisions---either periodically or by an event notification---it invokes the procedure of deriving switching decisions (see Figure~\ref{fig:Interrupt driven flowchart of switching coordinator}) as~follows:
\begin{enumerate}
    \item Initiate the \emph{\hl{get network status}} procedure, consisting of 
    retrieving the list of configurations
    and the current network state, i.e.,~operational data.
    \item Run the \emph{\hl{calculate configurations}} procedure, which executes the optimization algorithm specified by the network administrator. This provides the optimal switching decisions in the form of configuration-duration tuples (in the case of periodic recalculation) or a configuration with infinite duration (in the case of event-based recalculation).
    \item In the case of periodic operation, run the process \emph{\hl{schedule recalculation}} to set up the next trigger time for recalculation.
\end{enumerate}

The implementation of a new configuration is invoked either immediately after deriving the switching decisions or when the duration of the ongoing configuration ends.
The procedure of implementing a configuration is carried out through a series of consecutive instructions executed by the switching coordinator, described in the~following:
\begin{enumerate}
    \item From the selected configuration, identify the set of physical links to activate using their respective \texttt{physical\_link\_id} fields from the list of configurations.
    If a physical link is already active in the ongoing configuration, it is ignored, and its identifier is excluded.
    Similarly, the set of physical links to deactivate is generated by identifying the currently active physical links that are not part of the target configuration.
    \item For the physical links to deactivate, the~switching coordinator
    instructs the SDN controller to stop these physical links and release the allocated resources.
    The SDN controller identifies the resources using the attribute \texttt{resources} in the list of links and issues the necessary deactivate and release commands.
    \item In the case of light-path switching (Section~\ref{sec:Switched QKD}), the~SDN controller performs a joint reconfiguration of the light paths by instructing the network elements accordingly.
    For example, a~reconfiguration of the input-port--output-port association of an optical switch.
    \item For the physical links to activate, the~switching coordinator
    instructs the SDN controller to tune and align the necessary resources in order to activate these physical links.
    The resources that constitute a physical link are identified using the attribute \texttt{resources} in the list of links. 
\end{enumerate}

The design described above is based on the principle that only the switching coordinator knows the current and desired configurations. The~SDN controller executes atomic operations on the infrastructure in response to the requests of the switching~coordinator.

\section{Optimization~Algorithms\label{sec:Optimization algorithms}}

In the context of the switching coordinator, optimization algorithms are the core mechanism through which high-level objectives, such as preventing key depletion, prioritizing certain communication paths, or balancing key generation, are translated into actionable reconfiguration plans.
These algorithms consume the unified network model from\mbox{ Section~\ref{subsec:Formal network model}} and produce concrete outputs that drive the switching and link activation over time.
As a building block, this is the configuration-duration tuple calculation introduced in Section~\ref{subsec:Periodic and event-driven operation} and
depicted in Figure~\ref{fig:Interrupt driven flowchart of switching coordinator} as \emph{\hl{calculate configurations}}.

Rather than being hardcoded for a single purpose, the~framework supports multiple optimization functions that may be selected or combined depending on the policy defined by the network administrator.
The two example strategies described in this section aim to maximize the key consumption rates supported by the network while preventing key depletion.
Such depletion is a critical condition in which links exhaust their supply of cryptographic keys,
which would disrupt key-dependent applications, making them unable to fulfill their security functions.
The described algorithms build on the foundations of recent research efforts, such as the optimization problems presented in~\cite{kanellos,ocn}.

\subsection{Assumptions}
For the optimization examples, a~constant key generation rate per physical link and a constant consumption rate per abstract link are assumed.
These constant rates are a good approximation for the average behavior of a stable network.
Note, however, that the proposed framework allows for fluctuations, taking them into account through periodic recalculations as well as events that can trigger new recalculations when needed.
Because of this and the bounds on the link buffers, it is assumed that the system converges with a deterministic strategy for any initial filling of the link buffers to a periodic steady state.
The link-buffer fillings at the beginning and end of a period are identical in the steady state; the~key gain in a period is zero in this~equilibrium.

It is assumed that the link buffers are sufficiently large such that no link buffer depletes during a steady-state time evolution when a non-negative average key gain is calculated in Equation~\eqref{eq:average key gain}.
The required link-buffer sizes are proportional to the period, which is assumed to be given.
For the analysis, all parameters are normalized relative to this given steady-state period.

Although key generation is paused during switching, the time required to switch is considered to be negligible.
This assumption is justified on the grounds that very low switching frequencies (e.g., no more than once per hour) are permitted by sufficiently large link~buffers.

\subsection{Mapping Between Network Model and~Optimization}

The consumption rates and link-buffer fillings are taken from the network model fields $\texttt{consumption\_rate}$ and $\texttt{buffer\_fill\_level}$ (Listing \ref{lst:current_state}) and
put into vectors $\bs{c}\in\mathbb{R}^L_{> 0}$ and $\bs{b}\in\mathbb{N}^L_0$, respectively.

Matrix $\bs{G} \in \mathbb{R}^{L \times C}_{\ge 0}$ is defined to map configurations to per-link key generation rates.
It is constructed from the information in Listing~\ref{lst:configs}.
In a network with $L$ abstract links and $C$ configurations, $\bs{G}$ has one column per configuration and one row per abstract link.
Each column contains the key generation rates for the different links with a specific configuration.
For example, the~column for configuration $\mathcal{C}_1$ from Equation~\eqref{eqn:simple configs} has positive entries for links A1B1, A2B2, and~A3B3 and zeros for the other~links.

When $\bs{G}$ is multiplied from the right by a standard unit vector $\bs{e}_c$, the~resulting vector $\bs{G}\bs{e}_c \in \mathbb{R}^L_{\ge0}$ contains the key generation rates for each link under configuration $c$.
Assuming a sequence of configurations is activated one after another with individual durations, the average key-generation rates during the total duration can be found by multiplying $\bs{G}$ from the right with a vector of the relative durations for each configuration: $\bs{G}\bs{p} \in \mathbb{R}^L_{\ge0}$.
Each element of $\bs{p} \in \mathbb{R}^C$ must satisfy $0 \le \bs{p}[c] \le 1$, $\forall c$, and~the elements cannot sum up to more than one, $\sum_{c=1}^C p[c] \le 1$.

Vector $\bs{p}$ translates into the configuration-duration tuple list from Section~\ref{subsec:Periodic and event-driven operation}.
This list is formed from all configurations with positive entries in $\bs{p}$.
The associated durations are the relative durations in $\bs{p}$ times the given duration of the period.
Because of the assumption of a steady state and sufficiently large link buffers, the order of the list is irrelevant.
The event-based operation of the switching coordinator is a special case where only one configuration is selected with an infinite~duration.

\subsection{Supported-Consumption-Rate~Metric}

The average key gain is defined as the difference between the average key generation rates $\bs{G}\bs{p}$ and the key consumption rates $\bs{c}$:
\begin{equation}
\bs{G}\bs{p} - \bs{c} \in \mathbb{R}^L\label{eq:average key gain}.
\end{equation}
The consumption rate per link is strictly positive, even with no key consumer, due to the outdating of key material.

To facilitate the investigation, the~consumption rate vector $\bs{c} = k\bs{\eta}$ is split into the weight vector $\bs{\eta}$ and the scalar $k$ \cite{kanellos}.
This allows us to turn Equation~\eqref{eq:average key gain} into Equation~\eqref{eq:relative average key gain} with a uniform consumption rate.
Vector $\bs{g} \in \mathbb{R}^L$ contains the relative average key gain of each link, which is the average key gain individually normalized by the key consumption weight per link.
\begin{equation}
\bs{g} = \diag\left(\bs{\eta}\right)^{-1} \left(\bs{G}\bs{p} - k\bs{\eta}\right) = \bs{\varGamma}\bs{p} - k\bs{1}\label{eq:relative average key gain}.
\end{equation}
$\diag\left(\bs{\eta}\right)^{-1}$ is an $L\times L$ diagonal matrix with one over each link consumption weight as its diagonal elements,
$\bs{\varGamma} = \diag\left(\bs{\eta}\right)^{-1}\bs{G}$ is the normalized mapping of configurations to per-link key generation rates, and~$\bs{1}$ is the all-ones vector of appropriate~size.

The largest $k$, for~which no entry of $\bs{g}$ is negative,
\begin{align}
  k_{\text{supported}} = \min_l\, (\bs{\varGamma}\bs{p})[l]\label{eq:k supported},
\end{align}
can be regarded as the largest supported-consumption-rate metric~\cite{kanellos}.
From uniform consumption rates in the original problem, $\bs{\eta}=\bs{1}$, it follows that $\bs{\varGamma} = \bs{G}$, and
$k_{\text{supported}}$ is the largest supported consumption rate of the weakest link. %please ensure intended meaning retained.
All entries of $\bs{\varGamma}\bs{p}$ are non-negative by definition.
In the following, it is assumed that there exists a positive $k_{\text{supported}}>0$.

\subsection{Fill Most Critical Buffer (FMCB)\label{subsec:Fill most critical buffer (FMCB)}}

One of the simplest, yet very effective strategies implemented as an example for notification-driven operation of the switching coordinator is the fill-most-critical-buffer policy.
This strategy is designed to react to potential key depletion events by identifying the link buffer in the network whose filling is closest to depletion and triggering key generation to replenish it as a~priority.

For this strategy, the~switching coordinator is not configured with periodic recalculation interrupts.
When a link buffer falls below a threshold, the~link buffer notifies the SDN controller, which relays this notification to the switching coordinator.
Upon receiving such a notification, the~switching coordinator reconfigures the network into the configuration that yields the highest key generation rate for this~link.

\begin{figure}[H]
\includegraphics[width=0.96\textwidth]{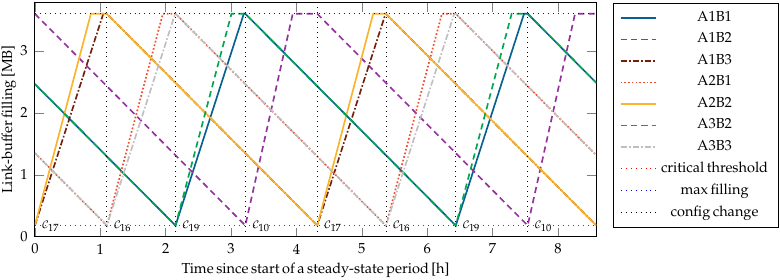}
  \caption{\hl{Steady-state} %MDPI: Please confirm whether an explanation of the c10, c17, c16, c19 needs to be added to the figure caption.
  %Authors: No explanation needed.
 time evolution with the FMCB strategy.
  Whenever the filling of a link buffer falls below the critical threshold, the network is reevaluated, and the configuration that recovers this link as fast as possible is activated.
  Links of transmitter $A_2$ are configured with the highest priority; links of $A_1$ are configured with the lowest priority.
  A uniform consumption rate close to \qty{25}{\percent} (\qty[per-mode = symbol]{2.35}{\kilo\bit\per\second}) of the lowest generation rate can be achieved for each link in the example setup, respectively.
  The evolution strongly depends on the network parameters, including the grace time (here \qty{600}{\second}) and maximum link-buffer fillings.
  \label{fig:FMCB_steady}}
\end{figure}

This strategy is reactive and single-goal driven: its main objective is to prevent imminent depletion of any individual link buffer.
Although it does not attempt to optimize global performance or anticipate long-term demand, it provides fast and deterministic responses to critical conditions, making it suitable for baseline operation or as a fallback strategy under constrained configurations.
A qualitative time evolution of the link-buffer fillings with FMBC is given in Figure~\ref{fig:FMCB_steady} for the network from Figure~\ref{fig:splitter network} with parameters defined in Section~\ref{subsec:Simulations}.

There is a grace time after a reconfiguration to mitigate inefficiencies when multiple links are close to depletion.
Notifications about critical buffers during the grace time will be handled jointly immediately after the grace time has passed.
In case multiple notifications about different link buffers have arrived during the grace time, the~switching coordinator will inquire the SDN controller after the grace time has passed about the current status of the~network.

Hereafter, with~$\bs{b}$ as the vector of current link-buffer fill levels for all links, the link whose buffer is closest to depletion will be identified as the most critical link:
\begin{eqnarray}
    l_{\mathrm{mc}} \in \argmin_{l}{\frac{\bs{b}[l]}{\bs{\eta}[l]}}\label{eq:CriticalLink}.
\end{eqnarray}

Note that it is very unlikely to have more than one link satisfying \eqref{eq:CriticalLink} because links have distinct operating parameters and rates. 
However, in~the case of obtaining more than one link satisfying \eqref{eq:CriticalLink}, either the link with the highest priority index is selected, or one is selected at random.
Links are assumed to have distinct priority indices predefined by the network~administrator.

Once a critical link $l_{\mathrm{mc}}$ is identified, the~configuration that maximizes the key generation rate for this link buffer is selected. % fill level of link $l_{\mathrm{mc}}$ at time $t^+$.
With standard unit vector $\bs{e}_{\mathrm{mc}}$, the~row with the available key generation rates for link $l_{\mathrm{mc}}$ is selected from the key generation mapping $\bs{\varGamma}$.
The configuration is selected that has the largest key generation rate for the critical link:
\begin{eqnarray}
  \bs{p}_{\mathrm{FMCB}} \in \argmax_{\bs{p} \in \left\{\bs{e}_1, ..., \bs{e}_C\right\}}\; \bs{e}_{\mathrm{mc}}^{\mathrm{T}}\bs{\varGamma}\bs{p},
\end{eqnarray}
where the solution $\bs{p}_{\mathrm{FMCB}}$ is a standard unit vector that only selects the optimal configuration for an infinite~duration.

In case multiple configurations provide the same optimal generation rate,
$\bs{\varGamma}$ can be reduced in a second iteration to only reflect the competing options, and the configuration that contributes to the second most-critical buffer among the possible links is~selected.

With linear consumption weights and generation rates, and because of the upper and lower bounds for the link buffers, the~FMCB strategy will eventually converge to a steady-state solution (see Figure~\ref{fig:FMCB_steady}).
In the steady-state solution, the~same series of configurations will be visited repeatedly in a loop, and the duration that the~network stays in a specific configuration will be the same in every repetition.
With this information, one can construct an average $\bs{p}$ and find the maximum supported-consumption-rate metric $k_{supported}$ from Equation~\eqref{eq:k supported} under this strategy.
The solution depends on the priorities of the different links, the~initial conditions, and~the grace~time.

\subsection{Maximize Minimal Average Key in the Link Buffers (MMAK)\label{subsec:Maximize minimal average key in the link buffers (MMAK)}}

The second strategy implemented in the switching coordinator aims to maintain long-term fairness in the network by ensuring that all key stores remain uniformly populated.
Unlike the FMCB strategy, which is reactive and focuses solely on the most depleted link buffer,
this approach seeks to equalize the average key levels across all link buffers, reducing the risk of depletion at any single point and improving the overall resilience of the~network.

This strategy operates proactively.
On a recalculation interrupt, the~switching coordinator evaluates the current network state information and
selects a list of network configurations and durations that balance the relative, average key gain across the network.
The switching coordinator then cycles through the list of configuration-duration tuples with config-change interrupts.
At every interrupt, the~network is reconfigured with the next configuration, and a new config-change interrupt is scheduled according to the duration.
As the key generation and consumption rates are constant for the simulations, it is not necessary to schedule another recalculation~interrupt.

The balancing is implemented by maximizing the lowest relative, average key gain in the network: %, $\min\left\{\bs{g}[l]\right\}$:
\begin{equation}
  \bs{p}_{MMAK} \in \argmax_{\bs{p} \ge \bs{0}}\;\min_l\left(\bs{g}[l]\right)\quad \mathrm{s.t.}\; \sum_c\bs{p}[c] \le 1
\end{equation}
or equivalently
\begin{equation}
  \bs{p}_{MMAK} \in \argmax_{\bs{p} \ge \bs{0}}\; k\quad \mathrm{s.t.}\quad \bs{1}^{\mathrm{T}}\bs{p} \le 1, \quad \bs{e}_l^{\mathrm{T}}\bs{\varGamma}\bs{p} \ge k\;\forall l,
\end{equation}
which is a direct optimization of the supported-consumption-rate metric from Equation~\eqref{eq:k supported}.

\subsection{Simulations}\label{subsec:Simulations}

\begin{figure}[H]
\includegraphics[width=0.97\textwidth]{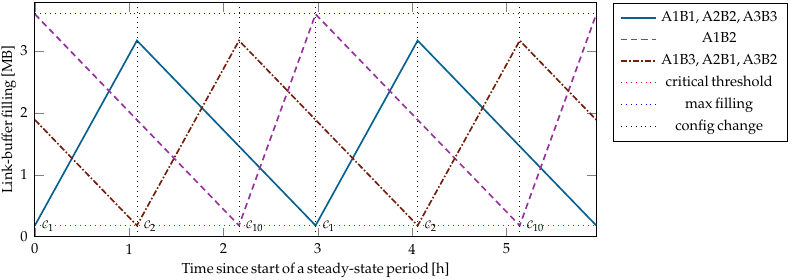}
  \caption{\hl{MMAK} %MDPI: 1. Please confirm whether an explanation of the c1, c2, c10 needs to be added to the figure caption. 2. We moved the figure after it was first mentioned in the main text. Please confirm.
  % Authors: 1. No explanation needed. 2. We confirm.
 returns the optimal configuration cycle $[\mathcal{C}_1, \mathcal{C}_2, \mathcal{C}_{10}]$ with relative durations for each configuration $[\frac{4}{11}, \frac{4}{11}, \frac{3}{11}]$.
  The largest possible period is selected, where all link buffers remain within the upper and lower bounds.
  Two periods of the steady-state time evolution are plotted.
  A uniform key consumption above \qty{36}{\percent} (\qty[per-mode = symbol]{3.49}{\kilo\bit\per\second}) of the lowest generation rate is supported in the given scenario, which is almost \qty{50}{\percent} higher than the supported rate with the FMCB strategy.
  \label{fig:MMAK_steady}}
\end{figure}

The proposed SDN-based network control, with the switching coordinator as an SDN application and the interface formalization from Section~\ref{subsec:Formal network model}, was implemented and tested for the example network in Figure~\ref{fig:splitter network}.
A time evolution of the link buffer fillings with the FMCB strategy, which allows for testing the notification-triggered recalculation of the network configuration, can be seen in Figure~\ref{fig:FMCB_steady}.
Whereas the evolution with MMAK, the~example strategy for cyclic configuration changes, is shown in Figure~\ref{fig:MMAK_steady}.

Although the numerical values of the parameters are arbitrarily selected, the~simulations give helpful insights into the qualitative behavior of the different strategies.
The key generation rates are set to be identical on all links:
$\qty[per-mode = symbol]{9.6}{\kilo\bit\per\second}$ without supporting receivers,
$\qty[per-mode = symbol]{11.2}{\kilo\bit\per\second}$ with a single supporting receiver,
and $\qty[per-mode = symbol]{12.8}{\kilo\bit\per\second}$ with two supporting receivers.
The capacity of each link buffer is set to \qty{3.6}{\mega\byte}, which corresponds to \qty{50}{\min} of key generation with the lowest generation rate.
The critical threshold, below~which a buffer is considered to be at risk of key starvation, 
is set to \qty{5}{\percent} of the maximum capacity (\qty{180}{\kilo\byte}) for all link buffers.
Uniform key consumption is assumed; all key consumption weights are set to one, $\bs{\eta} = \bs{1}$.

The simulations clearly demonstrate the strengths of the MMAK strategy with cyclic configuration changes in contrast to the reactive approach of FMCB.
By proactively maintaining a balanced distribution of key material across all buffers, MMAK optimally allocates the network resources to achieve the highest key throughput on the worst link.
For the example network, MMAK finds the smallest set of configurations necessary to fill all link buffers.
Only 3 out of the 22 \emph{\hl{useful}} configurations are selected, and the supported uniform key consumption is higher than \qty{36}{\percent} of the lowest key generation rate in the network.
Physical links with supporting receivers will only be activated when they benefit the\linebreak whole network.
MMAK has the disadvantage that its complexity grows prohibitively for larger networks. % \cite{kanellos}.

{\looseness=-1
In contrast, FMCB returns the best configuration that prevents depletion of the most critical buffer when triggered by a notification from the network.
But favoring the critical buffers might drive the network into a suboptimal evolution, since the configurations with supporting receivers are always favored.
Even with a careful selection of the link priorities, FMCB activates at least 4 different configurations with no more than two links active at a time.
The supported uniform key consumption is less than \qty{25}{\percent} of the lowest key generation rate in the network, which is less than \qty{70}{\percent} of the supported uniform key consumption with the MMAK strategy.
Despite the suboptimal performance in the given scenario, FMCB is still a reasonable strategy because of its simplicity.
If the example network were changed such that only transmitter $A_1$ allows the protocol with supporting receivers, and assuming a proper selection of the link priorities, FMCB would converge to the same result as~MMAK.
\par}

\section{Conclusions and Future~Work}\label{sec:Conclusions}

In this work, a~novel SDN-based framework for QKD networks is proposed that allows an efficient management of the switching capabilities of QKD modules.
In particular, the~switching coordinator is developed as an SDN application to continuously control the switching among QKD modules across the network.
Furthermore, a~formal network model is proposed to communicate necessary operational data, including the capabilities of QKD modules, real-time states of link buffers, and~real-time rates of key generation and key consumption on all network~links.

The association among QKD modules is updated by the switching coordinator in an interrupt-driven manner, either periodically and/or upon the reception of critical event notifications.
Whenever triggered, the~switching coordinator derives the next configuration, or~a set of configuration-duration tuples, 
based on switching strategies defined by the network administrator and the real-time network information provided in the SDN~controller.

The proposed framework is evaluated on an example switched QKD network according to two switching strategies: FMCB and MMAK.
Simulation results demonstrate that the MMAK strategy outperforms FMCB in maintaining a balanced distribution of key material across all link buffers, albeit with rapidly increasing complexity for larger networks.
In contrast, FMCB serves as a simple reactive approach that effectively alleviates critical conditions.
However, it may lead to suboptimal network evolution due to its localized~objective.

This work provides a reliable and flexible foundation for the deployment of SDN-enabled switching in QKD networks.
In ongoing efforts, the framework and switching coordinator are being deployed and tested in a real-world SDN-based QKD network environment.
This will allow us to refine the framework by investigating the impact of hardware restrictions, e.g.,~the time it takes to switch.
Future switching strategies could be based on policy-driven goals, such as prioritizing specific links, nodes, or~applications based on their strategic relevance.
This deployment will enable a better understanding of operational constraints and pave the way for the development of more robust, scalable, and~autonomous QKD management~systems.

%%%%%%%%%%%%%%%%%%%%%%%%%%%%%%%%%%%%%%%%%%
\vspace{6pt} 

%%%%%%%%%%%%%%%%%%%%%%%%%%%%%%%%%%%%%%%%%%

% Citation: Mendez, R. B.; Brunner, H. H.; Brito, J. P.; Taramit, H.; Fung, C.-H. F.; Pastor, A.; Cantó, R.; Folgueira, J.; Lopez, D. R.; Peev, M.; Martin, V.

% \authorcontributions{For research articles with several authors, a short paragraph specifying their individual contributions must be provided. The following statements should be used ``Conceptualization, X.X. and Y.Y.; methodology, X.X.; software, X.X.; validation, X.X., Y.Y. and Z.Z.; formal analysis, X.X.; investigation, X.X.; resources, X.X.; data curation, X.X.; writing---original draft preparation, X.X.; writing---review and editing, X.X.; visualization, X.X.; supervision, X.X.; project administration, X.X.; funding acquisition, Y.Y. All authors have read and agreed to the published version of the manuscript.'', please turn to the  \href{http://img.mdpi.org/data/contributor-role-instruction.pdf}{CRediT taxonomy} for the term explanation. Authorship must be limited to those who have contributed substantially to the work~reported.}

\authorcontributions{Conceptualization, R.B.M., H.H.B., C.-H.F.F. and J.P.B.; 
methodology, R.B.M., H.H.B., J.P.B., A.P. and R.C.; 
software, R.B.M. and H.H.B.; 
validation, R.B.M., H.H.B. and H.T.; 
formal\linebreak analysis, R.B.M., H.H.B., H.T., C.-H.F.F., J.P.B., A.P. and R.C.; 
investigation, M.P., V.M., D.R.L. and J.F.; 
resources, M.P., A.P., D.R.L., R.C. and J.F.; 
data curation, R.B.M.,  H.H.B., H.T. and C.-H.F.F.; 
writing---original draft preparation, R.B.M., H.H.B. and H.T.; 
writing---review and editing, R.B.M., H.H.B., H.T., C.-H.F.F., J.P.B. and M.P.; 
visualization, R.B.M., H.H.B. and H.T.; 
supervision, M.P. and V.M.; 
project administration, M.P. and V.M.; 
funding acquisition, V.M. 
All authors have read and agreed to the published version of the manuscript.}

\funding{\hl{This} %MDPI: Information regarding the funder and the funding number should be provided. Please check the accuracy of funding data and any other information carefully.
%Author: We have revised the funding and we updated it. The previous version is commented.
%work received funding from the «Hub Nacional de Excelencia en Comunicaciones Cuánticas»,  Ministerio para la Transformación Digital y de la Función Pública of Spain (TSI-100152-2025-1) and the European Union-NextGenerationEU; the MADQuantum-CM, funded by the Comunidad de Madrid (Programa de Acciones Complementarias); the Recovery, Transformation and Resilience Plan-Funded by the European Union-(NextGenerationEU, PRTR-C17.I1); and QUITEMAD, CAM Programa TEC-2024/COM-84~QUITEMAD-CM.
\hl{work received funding from projects “MADQuantum-CM”, funded by Comunidad de Madrid (Programa de Acciones Complementarias) and by the Recovery, Transformation and Resilience Plan—Funded by the European Union—NextGenerationEU, (PRTRC17.I1) and “Hub Nacional de Excelencia en Comunicaciones Cuánticas” project, funded by Ministerio para la Transformación Digital y de la Función Pública and by the Recovery, Transformation and Resilience Plan—Funded by the European Union—NextGenerationEU (PRTR-C16.R1).}
 }

\dataavailability{\hl{The original contributions presented in this study are included in the article. Further inquiries can be directed to the corresponding author.} %MDPI: We add Data Availability Statements. Please confirm.}. 
%Authors: We confirm.
} 

\conflictsofinterest{Author H.H.B., C.-H.F.F. and M.P. were employed by the company Huawei Technologies Duesseldorf GmbH. Author A.P., R.C., J.F. and D.R.L. were employed by the company Telefónica gCTIO. The remaining authors declare that the research was conducted in the absence of any commercial or financial relationships that could be construed as a potential conflict of interest. %MDPI: For commercial affiliations, all authors must be accounted for. We recommend using the following template: “Author XXXXXXX was employed by the company XXXXXXX. The remaining authors declare that the research was conducted in the absence of any commercial or financial relationships that could be construed as a potential conflict of interest.” Please check whether there are any potential commercial interests that need to be declared according to the relevant guidelines, which can be found at the following link: https://www.mdpi.com/ethics#_bookmark17.
%Author: We have no conflicts of interest. We have provided with our first submission "the completed, signed MDPI Disclosure Form".
} 

%%%%%%%%%%%%%%%%%%%%%%%%%%%%%%%%%%%%%%%%%%
%% Optional
\appendixtitles{no} % Leave argument "no" if all appendix headings stay EMPTY (then no dot is printed after "Appendix A"). If~the appendix sections contain a heading then change the argument to "yes".
\appendixstart
%\appendix
%\section[\appendixname~\thesection]{}
%\subsection[\appendixname~\thesubsection]{}
% The appendix is an optional section that can contain details and data supplemental to the main text---for example, explanations of experimental details that would disrupt the flow of the main text but nonetheless remain crucial to understanding and reproducing the research shown; figures of replicates for experiments of which representative data are shown in the main text can be added here if brief, or as Supplementary Data. Mathematical proofs of results not central to the paper can be added as an appendix.

% All appendix sections must be cited in the main text. In the appendices, Figures, Tables, etc. should be labeled, starting with ``A''---e.g., Figure A1, Figure A2, etc.

% \input{appendix}

%%%%%%%%%%%%%%%%%%%%%%%%%%%%%%%%%%%%%%%%%%
%\isPreprints{}{% This command is only used for ``preprints''.
\begin{adjustwidth}{-\extralength}{0cm}

\reftitle{References}



% If authors have biography, please use the format below
%\section*{Short Biography of Authors}
%\bio
%{\raisebox{-0.35cm}{\includegraphics[width=3.5cm,height=5.3cm,clip,keepaspectratio]{Definitions/author1.pdf}}}
%{\textbf{Firstname Lastname} Biography of first author}
%
%\bio
%{\raisebox{-0.35cm}{\includegraphics[width=3.5cm,height=5.3cm,clip,keepaspectratio]{Definitions/author2.jpg}}}
%{\textbf{Firstname Lastname} Biography of second author}

% For the MDPI journals use author-date citation, please follow the formatting guidelines on http://www.mdpi.com/authors/references
% To cite two works by the same author: \citeauthor{ref-journal-1a} (\citeyear{ref-journal-1a}, \citeyear{ref-journal-1b}). This produces: Whittaker (1967, 1975)
% To cite two works by the same author with specific pages: \citeauthor{ref-journal-3a} (\citeyear{ref-journal-3a}, p. 328; \citeyear{ref-journal-3b}, p.475). This produces: Wong (1999, p. 328; 2000, p. 475)

%%%%%%%%%%%%%%%%%%%%%%%%%%%%%%%%%%%%%%%%%%
\PublishersNote{}
%\isPreprints{}{% This command is only used for ``preprints''.
\end{adjustwidth}
\end{document}